\documentclass[12pt]{iopart}

\usepackage{graphicx}
\usepackage{cite}

\begin{document}

\title[On the number of contacts of crosslinked polymer chains on fractals]
{On the number of contacts of a floating polymer chain crosslinked
with a surface adsorbed chain on  fractal structures}

\author{Ivan \v Zivi\'c}

\address{ Faculty of Natural Sciences and
Mathematics, University of Kragujevac, 34000 Kragujevac, Serbia}

\eads{\mailto{} \mailto{ivanz@kg.ac.yu}}

\begin{abstract}
We study the interaction problem of a linear polymer chain,
floating in fractal containers that belong to the
three-dimensional  Sierpinski gasket (3D SG) family of fractals,
with a surface-adsorbed linear polymer chain. Each member of the
3D SG fractal family has a fractal impenetrable 2D adsorbing
surface, which appears to be 2D SG fractal.  The two-polymer
system is modelled by two mutually crossing self-avoiding walks.
By applying the Monte Carlo Renormalization Group (MCRG) method,
we calculate the critical exponents $\varphi$, associated with the
number of contacts of the 3D SG floating polymer chain, and the 2D
SG adsorbed polymer chain, for a sequence of SG fractals with
$2\le b\le 40$. Besides, we propose  the codimension additivity
(CA) argument formula for $\varphi$, and compare its predictions
with our reliable set of the  MCRG data. We find that $\varphi$
monotonically decreases with increasing $b$, that is, with
increase of the container fractal dimension. Finally, we discuss
the relations between different contact exponents, and analyze
their possible behaviour  in the fractal-to-Euclidean crossover
region $b\to\infty$.
\end{abstract}

\maketitle

\section{Introduction}
\label{uvod}

The self-avoiding walk (SAW) is a random walk that must not
contain self-intersections. This kind of random walk, placed on a
lattice, has been widely used as a model of a linear polymer in a
good solvent \cite{vc}. Even though an isolated polymer chain is
difficult to observe experimentally, numerous studies of the
single-chain statistics have been upheld as a requisite step
towards understanding the statistics of collection-chain systems
\cite{ow}. A reasonable  extension of a single polymer concept is
a model of two interacting linear polymers (two interacting SAWs),
which has been a subject of extensive studies because of its
theoretical and practical interest. To investigate  the critical
properties of the two-chain (or  many-chain) systems various
theoretical techniques have been applied, including the field
theoretical approach \cite{stara13}, Monte Carlo simulations
\cite{stara14,pelissetto}, transfer-matrix calculations
\cite{stara9}, and renormalization group (RG) methods
\cite{stara12,epj1,WS05}. Among other fields, a system of two
interacting SAWs, that are mutually avoiding, has been
successfully applied as a model of diblock copolymers
\cite{stella1,stella2}, as well as in the studies of
double-stranded  DNA molecules \cite{dna1,dna2,dna3}.

In order to study the number of contacts between monomers that
belong to different polymers, the two-polymer system may  be
modelled by two mutually crossing self-avoiding walks
\cite{ks93,leoni}, that is, by two SAWs whose paths on an
underlying lattice can cross (intersect) each other. Each crossing
between two SAW paths corresponds to a contact of two monomers
that belong to different polymer chains, and therefore with each
crossing we may associate the contact energy $\epsilon_c$. In
analogy with the problem of polymer adsorption onto a hyperplane
\cite{debel}, we may assume that with decreasing of temperature
the number of crossings $M$ increases so that at the critical
temperature $T_c$ it behaves according to the power law
\begin{equation}\label{j1}
   M\sim N^\varphi\>,
\end{equation}
where $N$ is the total number of monomers in the longer chain, and
$\varphi$ is the crossover critical exponent. Below $T_c$ the
number of contacts becomes proportional to $N$, whereas above
$T_c$ it is vanishingly small.

In spite of  different approaches to the two-SAW problem, the
entire physical picture achieved so far, is almost entirely of a
phenomenological character and there are only few numerical
results. For instance, in the case of Euclidean lattices, besides
the codimension additivity (CA) argument predictions for the
contact critical exponents (with conjectures $\varphi=1/2$ in
$d=2$,  and $\varphi\approx1/5$ in $d=3$), there is only Monte
Carlo  result $\varphi= 0.516\pm0.005$, for a model of two
mutually crossing SAWs on the square lattice \cite{leoni}. The
above problem has been also studied on fractal lattices. But, in
the case of fractals it may be noted that the two-dimensional (2D)
fractal lattices (embedded in the 2D Euclidean space) have been
more frequently investigated \cite{ks93,z11,z17} than the
corresponding model in the case of  3D fractal structures
\cite{ksjstat}. Since the 3D fractals definitely may serve as a
better description of real systems (porous media, for instance),
it is recommendable to study the described model in a case of a
set of 3D fractals. For this reason, it is desirable to extend the
study  of two-polymer system on a family of fractal lattices whose
characteristics approach properties of a 3D Euclidean lattice.

In this paper we report results of our study for the contact
critical exponent $\varphi$ of two polymer chains, that display
inter-chain interactions, in the three-dimensional  fractal
lattices that belong to the Sierpinski gasket (SG) family of
fractals. The two-polymer system is modelled by two SAWs which are
allowed to cross each other. We assume that the first polymer is a
floating chain in the bulk of 3D SG fractal, while the second
polymer chain is adsorbed onto one of the four surfaces (which is,
in fact, 2D SG fractal) by which is 3D SG fractal bounded. In
section~\ref{druga} of the paper, we first describe the 3D SG
fractals for general $b$. Then, we present the framework of the RG
method for studying statistics of two mutually crossing SAW
chains, taking into account the presence of the inter-chain
interactions, in a way that should make the method transparent for
the Monte Carlo  calculations of the contact critical exponent
$\varphi$. In section~\ref{treca} we present the obtained specific
values of $\varphi$ for a sequence of 3D SG fractals, that is, for
$2\le b\le 40$. In the same section we propose the
phenomenological formulae for $\varphi$, based on the CA
arguments,  test their predictions on the obtained MCRG data, and
discuss  relations between different contact exponents. Summary of
the obtained results and the relevant conclusions are given in
section~\ref{cetiri}.

\section{Monte Carlo renormalization group approach}
\label{druga}

In this section we are going to apply the Monte Carlo
renormalization group (MCRG) method to the studied  model of
interacting polymers on the 3D SG family of fractals. These
fractals have been studied in numerous papers so far, and
consequently we shall give here only a requisite brief account of
their basic properties. It starts with recalling the fact that
each member of the 3D SG fractal family is labelled by an integer
$b\ge2$ and can be constructed in stages.  At the first stage
($r=1$) of the construction there is a tetrahedron of base $b$
that contains  $b(b+1)(b+2)/6$ upward oriented unit tetrahedrons.
The subsequent fractal stages are constructed recursively, so that
the complete self-similar fractal lattice can be obtained as the
result of an infinite iterative process of successive $(r\to r+1)$
enlarging the fractal structure $b$ times, and replacing the
smallest parts of enlarged structure with the initial structure
$r=1$ (see for instance, figure~1 of \cite{z15}). Fractal
dimension $d_f$ of the 3D SG fractal is equal to
\begin{equation}\label{fd3dsg}
d_f^{3D}={{\ln [{{b(b+1)(b+2)}/6}}]/{\ln b}}\>.
\end{equation}
We assume here that one of the four boundaries of the 3D SG
fractal is an impenetrable  adsorbing surface, which is itself a
2D SG fractal with the fractal dimension
\begin{equation}\label{fd2dsg}
d_f^{2D}=\ln[b(b+1)/2]/\ln b\>.
\end{equation}

\begin{figure}
\hskip5cm
\includegraphics[scale=0.3]{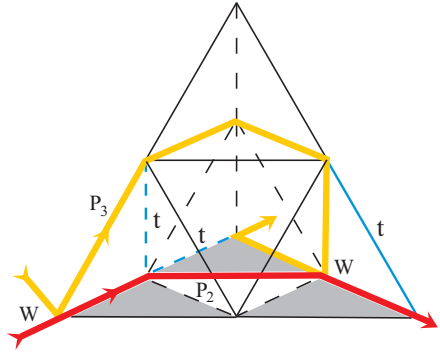}
\caption{The structure of the three-dimensional SG fractal, for
$b=2$, at the first stage of construction, with an example of the
bulk polymer chain ($P_3$) depicted by yellow  line and the
surface-adsorbed polymer chain ($P_2$) depicted by red line. The
shaded area represents the adsorbing surface (the two-dimensional
SG fractal). The two polymers are crosslinked at the two cites, so
that each contact contributes the weight factors $w$. The blue
bonds, marked by $t$, describe the interactions between those
monomers which are nearest neighbors to the crosslinked points.
Thus, the depicted two-polymer configuration contributes the
weight $x_3^{5}x_2^{3}w^2t^3$ in the corresponding RG equation
(especially, in the equation (\ref{eq:RGAi}) for $i=2$ and
$r=0$).}
 \label{fig1}
\end{figure}

In the terminology that applies to the SAW, we assign the weight
$x_3$ to a step of the SAW in the bulk (3D SG fractal), which
represents a floating polymer (we mark it by $P_3$), and the
weight $x_2$ to a step of the SAW adsorbed on the surface (2D SG
fractal), which represents an adsorbed polymer (marked by $P_2$),
whose monomers act as pining cites for the floating polymer.  In
order to explore interacting effects of two SAWs on the underlying
fractals, we introduce the two Boltzmann factors
$w=e^{-\epsilon_c/k_BT}$ and $t=e^{-\epsilon_t/k_BT}$, where
$\epsilon_c$ is the energy of two monomers in contact (which
occurs at a crossing site of SAWs), while $\epsilon_t$ is the
energy associated with two sites which are nearest neighbours to a
crrosslinked site and which are visited by different SAWs (see
figure \ref{fig1}).

\begin{figure}
\hskip5cm
\includegraphics[scale=0.4]{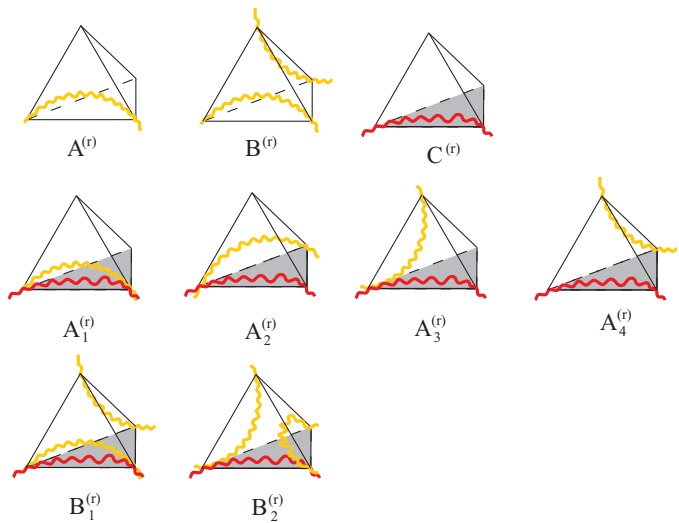}
\caption{ \label{fig2} Schematic representation of the nine
restricted generating functions used in the description of all
possible two-SAW configurations, within the $r$-th stage of the 3D
SG fractal structure. Thus, for  example, the $A_1^{(r)}$
represents the configuration when both polymers (3D SG floating
and 2D SG adsorbed) pass through the same vertices of a particular
$r$-th stage of the fractal lattice. The interior details of the
$r$-th stage fractal structure, as well as  details of the chains,
are not shown (for the chains, they are manifested by the wiggles
of the SAW paths).}
\end{figure}

To describe all possible configurations of the two-chain polymer
system, within the accepted  model, we need to introduce the nine
restricted partition functions $A^{(r)}$, $B^{(r)}$, $C^{(r)}$,
$A_i^{(r)}\, (i=1,2,3,4)$, and $B_i^{(r)}\, (i=1,2)$, that are
depicted in figure \ref{fig2}. The recursive nature of the fractal
construction implies the following recursion relations for
restricted partition functions
\begin{eqnarray}
\fl A^{(r+1)}=\sum_{N_{A},N_{B}} a(N_{A},N_{B})\, A^{N_{A}}
B^{N_{B}}\,,
\label{eq:RGA}\\
\fl B^{(r+1)}=\sum_{N_{A},N_{B}} b(N_{A},N_{B})\, A^{N_{A}}
B^{N_{B}}\,,
\label{eq:RGB}\\
\fl C^{(r+1)}=\sum_{N_C} c(N_C)\, C^{N_C}\,,
\label{eq:RGC}\\
\fl  A_i^{(r+1)}=  \sum_{\cal{N}}
     a_i({\cal{N}})\,
 A^{N_A}B^{N_B}C^{N_C}
\prod_{j=1}^{4} A_{j}^{N_{A_j}}
\prod_{k=1}^2B_{k}^{N_{B_k}}\,,\quad i=1,2,3,4\>,
\label{eq:RGAi}\\
\fl  B_i^{(r+1)}=  \sum_{\cal{N}} b_i({\cal{N}})\,
 A^{N_A}B^{N_B}C^{N_C}
\prod_{j=1}^{4} A_{j}^{N_{A_j}}\prod_{k=1}^2
B_{k}^{N_{B_k}}\,,\quad i=1,2\>, \label{eq:RGBi}
\end{eqnarray}
where $\cal{N}$ denotes the set of numbers
${\cal{N}}=\{N_{A},N_{B},N_{C},N_{A_1},N_{A_2},N_{A_3},N_{A_4},
N_{B_1},N_{B_2}\}$, and, where we have omitted the superscript
$(r)$ on the right-hand side of the above relations. The
self-similarity of the fractals implies that the sets of
coefficients $a,b$ and $c$, that describe a single polymer
configurations ($a$ and $b$ coefficients describe the bulk polymer
configurations, whereas the $c$ coefficients describe the adsorbed
polymer configurations), and the sets of coefficients $a_i$ and
$b_i$ of the corresponding two-chain configurations do not depend
on $r$. Each of the two-chain coefficients ($a_i$ and $b_i$)
represents the number of ways in which the corresponding parts of
the two-SAW configuration, within the $(r+1)$-th stage fractal
structure, can be comprised of the two-SAW configurations within
the fractal structures of the next lower order. Because of the
independence of $r$, these coefficients can be calculated by
studying all two-SAW paths within the fractal generator only, that
is on the first step of construction $r=1$.

The above set of relations (\ref{eq:RGA})--(\ref{eq:RGBi}), can be
considered as the RG equations for the problem under study, with
the  initial conditions: $A^{(0)}=x_3$, $B^{(0)}=x_3^2$,
$C^{(0)}=x_2$,  $A_1^{(0)}=x_3x_2w^2$,
$A_2^{(0)}=A_3^{(0)}=x_3x_2wt$, $A_4^{(0)}=x_3x_2$ and
$B_1^{(0)}=B_2^{(0)}=x_3^2x_2w^2$, which correspond to the unit
tetrahedron. On the physical grounds that are reasonable for the
studied model \cite{kspa1}, one can expect that the RG
transformations should have three relevant fixed points of the
general type
\begin{eqnarray}\label{fp}
 (A^*,B^*,C^*,A_1^*,A_2^*,A_3^*,A_4^*,B_1^*,B_2^*).
\end{eqnarray}
The first fixed point
\begin{eqnarray}\label{fp1}
 (A^*,B^*,C^*,0,0,0,A_4^*,0,0),
\end{eqnarray}
with $A_i^*=0\,(i=1,2,3)$, $A_4^*=A_4^*(A^*,B^*,C^*)$ and
$B_i^*=0\,(i=1,2)$, due to the meaning of these quantities (see
figure 2), describes segregated phase of two chain polymers that
should be expected in the high temperature region $T>T_c$. The
second fixed point
\begin{eqnarray}\label{fp2}
 (A^*,B^*,C^*,A^*C^*,A^*C^*,A^*C^*,A^*C^*,B^*C^*,B^*C^*),
\end{eqnarray}
with $A_i^*=A^*C^*\,(i=1,2,3,4)$ and $B_i^*=B^*C^*\,(i=1,2)$,
which appears to be a tricritical point, describes the state of
the two-polymer system that occurs at the critical temperature
$T=T_c$ when segregated and entangled polymer phases become
identical. Finally, the third fixed point
\begin{eqnarray}\label{fp3}
 (0,0,0,C^*,0,0,0,0,0),
\end{eqnarray}
with the nonzero value only for $A_1^*=C^*$, describes the polymer
entangled state, which should appear at low temperatures $T<T_c$.

In what follows we focus our attention on the tricritical fixed
point (\ref{fp2}) to calculate the contact critical exponent
$\varphi_{32}$ between $P_3$ and $P_2$ polymer chains. We can
observe that the above system of RG equations
(\ref{eq:RGA})--(\ref{eq:RGBi}) can be split   into three
uncoupled sets of RG equations: (\ref{eq:RGA})--(\ref{eq:RGB}),
(\ref{eq:RGC}), and (\ref{eq:RGAi})--(\ref{eq:RGBi}). Moreover, it
should be noticed, that for each $b$, the first two sets of RG
equations, (\ref{eq:RGA})--(\ref{eq:RGB}) and (\ref{eq:RGC}), have
only one nontrivial fixed point value $(A^*,B^*)$ \cite{z15} and
$C^*$ \cite{EKM} respectively, which thereby completely determine
the coordinates of the tricritical fixed point. Calculation of
$\varphi_{32}$ starts with solving the eigenvalue problem of the
RG equations (\ref{eq:RGA})--(\ref{eq:RGBi}), linearized at the
tricritical fixed point. The related eigenvalue problem can be
separated into three parts. The first part of the eigenvalue
problem, related to the equations (\ref{eq:RGA})--(\ref{eq:RGB})
gives the eigenvalue $\lambda_{\nu_3}$ of the end--to--end
distance critical exponent $\nu_3=\ln b/\ln\lambda_{\nu_3}$ of the
3D SG floating polymer, while the second part, related to the
equation (\ref{eq:RGC}), gives the eigenvalue $\lambda_{\nu_2}$ of
the end--to--end distance critical exponent $\nu_2=\ln
b/\ln\lambda_{\nu_2}$ of the 2D SG SAW, that represents the
adsorbed polymer. Finally, the third part of the eigenvalue
problem (related to the equations
(\ref{eq:RGAi})--(\ref{eq:RGBi})) reduces to solving the equation
\begin{equation}\label{det15}
    \mbox{det}\left| \left({\partial X'_i\over \partial X_j}
    \right)^{*}-
    \lambda\,\delta_{ij} \right|=0,
\end{equation}
where $X_i$ are elements of the set $\{A_1,A_2,A_3,A_4,B_1,B_2\}$,
and the asterisk means that the derivatives should be taken at the
tricritical fixed point. Also, we have used the prime symbol as a
superscript for the $(r+1)$--th restricted partition functions and
no indices for the $r$--th order partition functions. The largest
eigenvalue $\lambda_{\varphi_{32}}$ of above equation determines
the contact critical exponent via the formula
\begin{equation}\label{fi}
  \varphi_{32}={\ln{\lambda_{\varphi_{32}}}\over \ln\lambda_{\nu_3}}.
\end{equation}

Hence, in an exact RG evaluation of $\varphi_{32}$ one needs to
calculate partial derivatives of sums
(\ref{eq:RGA})--(\ref{eq:RGBi}), and thereby one should find the
coefficients $a,b,c,a_i$, and $b_i$. The latter can be calculated
by an exact enumeration of all possible two-SAW configurations for
each particular $b$. We have found that this enumeration is
feasible only for small $b$. However, for large $b$ the exact
enumeration turns out to be a forbidding task. We  bypass this
problem by applying the MCRG method. Within this method, the first
step would be to locate the tricritical fixed point. To this end,
we may observe that the results obtained in \cite{z15} provide
information for both $(A^*,B^*)$ and $\lambda_{\nu_3}$ for a
sequence of 3D SG fractals with $2\le b\le 40$, whereas the
results obtained in \cite{z1} give the fixed point values of $C^*$
for 2D SG fractals, for $b$ in the range $2\le b\le 80$ .
Accordingly, the next step in the MCRG method consists of finding
$\lambda_{\varphi_{32}}$ without explicit calculation of the RG
equation coefficients.

To solve the partial eigenvalue problem (\ref{det15}), so as to
learn $\lambda_{\varphi_{32}}$, we need to find the requisite 36
partial derivatives. These derivatives can be related to various
averages of the numbers $N_{A_i}$ and $N_{B_i}$ of different
crossings of the SAWs for various two--SAW  configurations that
correspond to the restricted partition functions $A_i^{(r)}$ and
$B_i^{(r)}$. For instance, starting with (\ref{eq:RGAi}) (in the
notation that does not use the superscripts $(r+1)$ and $r$) and
by differentiating it with respect to $A_1$ we get
\begin{eqnarray}
\fl{\partial{A_i'}\over\partial{A_1}}= \sum_{\cal{N}}
     N_{A_{1}}a_i({\cal{N}})\,
 A^{N_A}B^{N_B}C^{N_C}A_{1}^{-1}
\prod_{j=1}^{4} A_{j}^{N_{A_j}}
\prod_{k=1}^2B_{k}^{N_{B_k}}\,,\quad i=1,2,3,4\>. \label{s1}
\end{eqnarray}
Now, assuming that $A_i'$ represents the grand canonical partition
functions for the ensemble of all possible  two-SAW
configurations, where each of two SAWs starts and leaves fractal
generator at two fixed corners, so that the first SAW is a 3D SG
floating chain, while the other SAW is a 2D SG adsorbed chain.
With this concept in mind, we can write the corresponding ensemble
average
\begin{eqnarray}
\fl{\langle N_{A_1}\rangle}_{A_i'}={1\over A_i'} \sum_{\cal{N}}
     N_{A_{1}}a_i({\cal{N}})\,
 A^{N_A}B^{N_B}C^{N_C}
\prod_{j=1}^{4} A_{j}^{N_{A_j}}
\prod_{k=1}^2B_{k}^{N_{B_k}}\,,\quad i=1,2,3,4\>, \label{s2}
\end{eqnarray}
which can be directly measured in a Monte Carlo simulation.
Combing (\ref{s1}) and (\ref{s2}) we can express the requisite
partial derivative in terms of the measurable quantity
\begin{equation}\label{s3a}
{\partial{A_i'}\over\partial{A_1}}={A_i'\over A_1}{\langle
N_{A_1}\rangle}_{A_i'}\,,\quad i=1,2,3,4\>.
\end{equation}
In a similar way we can find the rest  derivatives, so that all
needed derivatives, calculated in the tricritical fixed point, may
be related with the corresponding averages
\begin{equation}\label{s3}
\fl{\left({\partial{A_i'}\over\partial{A_j}}\right)}^*={\langle
N_{A_j}\rangle}_{A_i'}^*\>,\quad
{\left({\partial{A_i'}\over\partial{B_k}}\right)}^*={A^*\over
B^*}{\langle N_{B_k}\rangle}_{A_i'}^*\,,\quad
i,j=1,2,3,4\>;\>\>k=1,2\>,
\end{equation}
\begin{equation}\label{s4}
\fl{\left({\partial{B_l'}\over\partial{A_j}}\right)}^*={B^*\over
A^*}{\langle N_{A_j}\rangle}_{B_l'}^*\>,\quad
{\left({\partial{B_l'}\over\partial{B_k}}\right)}^*={\langle
N_{B_k}\rangle}_{B_l'}^*\,,\quad j=1,2,3,4\>;\>\>k,l=1,2\>.
\end{equation}
Consequently to solve the  eigenvalue problem (\ref{det15}), so as
to learn $\lambda_{\varphi_{32}}$, we need to find the above
partial derivatives at the tricritical fixed point. These
derivatives are related to various averages of the numbers
$N_{A_j}$, and $N_{B_k}$,  of different two-SAW parts (of the
types $A_j$, and $B_k$) within the corresponding  two-SAW
configurations (described with $A'_i$ or $B'_l$). Therefore, to
calculate the derivatives (\ref{s3})--(\ref{s4}) at the
tricritical fixed point, one needs 36 averages ($\langle
N_{A_j}\rangle_{A'_i}^*$, $\langle N_{B_k}\rangle_{A'_i}^*$,
$\langle N_{A_j}\rangle_{B'_l}^*$, $\langle
N_{B_k}\rangle_{B'_l}^*$), which are all measurable through Monte
Carlo simulations.  The pertinent Monte Carlo technique has been
described  in \cite{z1}, and we would not like to elaborate on
them here. Solving numerically the eigenvalue equation
(\ref{det15}) we obtain $\lambda_{\varphi_{32}}$, and, finally,
using relation ($\ref{fi}$), we calculate the contact critical
exponent $\varphi_{32}$, whose  specific values we present in the
next section.

\section{Results and discussion}
\label{treca}

We have studied the interaction problem of a floating polymer
chain confined in the 3D SG fractal container, with a surface (2D
SG fractal) adsorbed polymer chain to calculate the  critical
exponent $\varphi_{32}$ which governs the number of contacts
between two chains.

The main goal of this study is the  MCRG evaluation of
$\varphi_{32}$, for various values of $b$. Furthermore, in the
case $b=2$, we have calculated the exact value. To find the exact
forms of RG equations (\ref{eq:RGA})--(\ref{eq:RGBi}), for $b=2$
fractal, using the computer facilities we have  been able to
enumerate all coefficients (which are available upon request
addressed to the author) that appear in these equations.
Linearization of obtained RG equations around the tricritical
fixed point (\ref{fp2}), in this case determined by the values
$A^*=0.4294$, $B^*=0.0499$ and $C^*=0.6180$ \cite{Dhar78}, gives
the following eigenvalues: $\lambda_{\nu_3}(b=2)=2.7965$ and
$\lambda_{\varphi_{32}}(b=2)=1.7475$, whereupon we  find the exact
value $\varphi_{32}(b=2)=\ln1.7475/\ln2.7965 =0.5428$. Here, we
notice that the same model is studied  on the four-simplex
lattice, which belongs to the same universality class as the $b=2$
3D SG fractal. In this study \cite{kspa1}  the  value
$\varphi_{32}=0.5669$ is reported, which appears as an approximate
result (and deviates 4\% from our exact finding), since in
approach applied in \cite{kspa1} some two-SAW configurations are
not taken into account (more precisely, the configurations
described by  function $A_2^{(r)}$, in our notation).
\Table{\label{tabela1} The MCRG ($2\le b\le 40$) results obtained
in this work for  the contact critical exponents $\varphi_{32}$
for the 3D SG family of fractals. For the completeness we quote
here the MCRG values for the RG parameters $(A^*,B^*)$ \cite{z15},
and $C^*$ \cite{z1}.}
\begin{tabular}{@{}lllll} \br
$b$ & $A^*$ & $B^*$ & $C^*$ &  $\varphi_{32}$\\
\mr
 2& 0.4311$\pm$0.0009& 0.0505$\pm$0.0023&
  0.61825$\pm$0.00061 & 0.5440$\pm$0.0056 \\
  3 & 0.3421$\pm$0.0004 &0.0245$\pm$0.0015& 0.55137$\pm$0.00044
&0.4969$\pm$0.0024 \\
  4 &  0.2898$\pm$0.0004& 0.0122$\pm$0.0020& 0.50658$\pm$0.00034
&0.4658$\pm$0.0006 \\
  5 &  0.2560$\pm$0.0004& 0.0067$\pm$0.0019& 0.47455$\pm$0.00028
 &0.4451$\pm$0.0012 \\
  6 &  0.2319$\pm$0.0003& 0.0038$\pm$0.0012& 0.45091$\pm$0.00024
&0.4250$\pm$0.0004\\
  7 &  0.2148$\pm$0.0003& 0.0020$\pm$0.0018& 0.43240$\pm$0.00021
 &0.4092$\pm$0.0008\\
  8 &  0.2016$\pm$0.0003& 0.0012$\pm$0.0026& 0.41780$\pm$0.00019
 &0.3963$\pm$0.0006\\
  9 &  0.1912$\pm$0.0004& 0.0007$\pm$0.0008& 0.40574$\pm$0.00017
  &0.3841$\pm$0.0007 \\
  10 &  0.1829$\pm$0.0003& 0.0005$\pm$0.0023& 0.39586$\pm$0.00007
& 0.3714$\pm$0.0005 \\
  12 &  0.1703$\pm$0.0004& 0.0001$\pm$0.0035& 0.38037$\pm$0.00013
 &0.3514$\pm$0.0004 \\
  15 &  0.1581$\pm$0.0001& -- & 0.36396$\pm$0.00011
 &0.3226$\pm$0.0003 \\
  17 &  0.1526$\pm$0.0001& -- & 0.35593$\pm$0.00008  &
  0.3126$\pm$0.0003 \\
  20 &  0.1462$\pm$0.0001& -- & 0.34681$\pm$0.00006  &
  0.2956$\pm$0.0003 \\
  25 &  0.1399$\pm$0.0001& -- &
0.33602$\pm$0.00008 &
  0.2677$\pm$0.0003\\
  30 &  0.1353$\pm$0.0001& -- &
   0.32876$\pm$0.00007&
  0.2573$\pm$0.0002 \\
  35 &  0.1327$\pm$0.0001& -- &
 0.32350$\pm$0.00008  &
  0.2183$\pm$0.0003 \\
  40 &  0.1305$\pm$0.0001& -- &
 0.31936$\pm$0.00006  &
  0.2016$\pm$0.0003\\
 \br
\end{tabular}
\endTable

For larger $b$, that is, for fractals in the range $2\le b\le 40$,
we have applied the MCRG method expounded in the previous section.
The obtained MCRG values for $\varphi_{32}$, together with the
pertaining error bars (determined from  statistics of measured
quantities through  the Monte Carlo simulations), are given in
table~\ref{tabela1}. The values of $\varphi_{32}$ for fractals
$2\le b\le 30$, have been calculated from averages obtained from
$10^6$ Monte Carlo simulations performed for each possible two-SAW
configuration, while $\varphi_{32}$ for fractals $b=35$ and $b=40$
was obtained performing $5\, 10^5$ Monte Carlo simulations. The
time needed to calculate all requisite averages ($\langle
N_{A_j}\rangle_{A'_i}^*$, $\langle N_{B_k}\rangle_{A'_i}^*$,
$\langle N_{A_j}\rangle_{B'_l}^*$, $\langle
N_{B_k}\rangle_{B'_l}^*$) increases exponentially with the scale
parameter $b$, so that for $5\, 10^5$ simulations performed on
fractals $b=10$ and $b=40$, at the corresponding tricritical fixed
points, it was required 10 minutes and 90 hours of the available
CPU time, respectively (on a PC with the Intel Pentium 4
processor).

\begin{figure}
\hskip3cm
\includegraphics[scale=0.4]{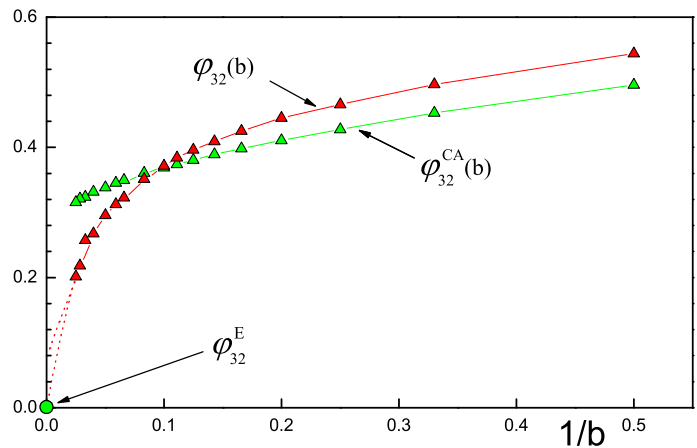}
\caption{Results obtained in this work for the contact critical
exponent $\varphi_{32}$ for the two-polymer system, when the first
polymer is a floating 3D SG chain, while the other is a 2D SG
adsorbed chain. The red triangles represent the MCRG values, while
green triangles correspond to the CA values calculated from
(\ref{foph}). The green circle depicts the tree-dimensional
Euclidean  value $\varphi_{32}^E\approx0$. The thin solid lines
(red and green) represent a simple interpolation of  data, while
the red dotted lines show the two possible scenarios for the
fractal-to-Euclidean behaviour of $\varphi_{32}$, and serve as the
guides to the eye. The error bars, for the MCRG data, are not
depicted in the figure, since in all cases they lie within of the
corresponding symbols (red triangles).} \label{fig3}
\end{figure}

Comparing our MCRG and exact result for $b=2$ fractal ($0.5440\pm
0.0056$ versus $0.5428$) we can see that the MCRG result deviates
0.22\% from the exact RG finding, which is a very  well agreement
between the two (Monte Carlo and exact) approaches of solving the
problem, and thereby provides reliance on accuracy of the MCRG
results for larger $b$. In figure \ref{fig3} we depict our MCRG
findings for the critical exponent $\varphi_{32}$, of the 3D SG
family of fractals, as a function of $1/b$. We  observe that the
critical exponent $\varphi_{32}$, in the region $2\le b\le40$, is
a monotonically decreasing function of $b$, which means that the
number of  polymer contacts (crossings of the SAW paths) decreases
with increasing of the fractal dimension of the SG fractals.

\Table{\label{tabela2} Numerical values  for the contact critical
exponents $\varphi_{32}$, $\varphi_{33}$ and $\varphi_{22}$
evaluated from the proposed phenomenological  formulae (based on
the CA arguments) (\ref{foph})--(\ref{ca22}). The values for
$\nu_3$ and $\nu_2$ have been calculated through the MCRG
simulations performed in \cite{z15} and \cite{z1}, respectively.}
\begin{tabular}{@{}rlllll} \br
$b\ $ &  $\nu_3$ &$\nu_2$ & $\varphi_{32}^{_{CA}}$&
$\varphi_{33}^{_{CA}}$&$\varphi_{22}^{_{CA}}$\\
\mr
  2& 0.6742&0.7985&0.4959&0.6516&0.7344\\
  3 & 0.6543&0.7937&0.4530&0.6287&0.7055\\
  4 & 0.6414 &0.7882 &0.4277&0.6140&0.6908\\
  5 & 0.6315 &0.7840 &0.4105&0.6050& 0.6808\\
 6 & 0.6239 &0.7801 &0.3981&0.5984&0.6745\\
  7 & 0.6169 &0.7773 &0.3890&0.5953&0.6689\\
  8 & 0.6130 &0.7742 &0.3805&0.5887&0.6658\\
  9 & 0.6087 &0.7722 &0.3738&0.5855&0.6622\\
  10 & 0.6048 &0.7698 &0.3690&0.5833&0.6603\\
12 & 0.5987 &0.7659 &0.3608&0.5792&0.6571  \\
 15 & 0.5933 &0.7620 &0.3497&0.5711&0.6529 \\
 17 & 0.5899 &0.7590 &0.3455 &0.5683 &0.6523 \\
20 & 0.5869 &0.7560 &0.3384 &0.5621 &0.6506\\
25 & 0.5817 &0.7516 &0.3317  &0.5577 &0.6495\\
30 & 0.5795 &0.7481 &0.3249  &0.5502 &0.6491 \\
35 &  0.5759 &0.7457 &0.3212 &0.5490 &0.6480\\
 40 & 0.5755 &0.7434 &0.3158 &0.5416 &0.6479\\
 \br
\end{tabular}
\endTable

In  figure~\ref{fig3}, we have also presented the numerical values
for $\varphi_{32}$, that follow from the  CA argument predictions
adopted for the studied model. Namely, the CA argument claims that
a codimension $d-d_{ip}$ of intersection points is a sum of
codimensions $d-d_1$ and $d-d_2$ of intersecting objects
\cite{leoni,Mand}.  Here, $d$ is a dimension of embedded space,
$d_{ip}$ dimension of intersection points, and, $d_1$ and $d_2$
dimensions of intersecting objects. In the studied problem the
embedded space is a 3D SG fractal of fractal dimension $d_f^{3D}$,
while the intersecting objects are linear polymers with the
fractal dimensions $d_1=1/\nu_3$ and $d_2=1/\nu_2$, whereupon
follows
\begin{equation}\label{caa1}
d_{ip}=\left({1\over\nu_3}+{1\over\nu_2}\right)-d_f^{3D}\>.
\end{equation}
On the other hand, the total number of intersection  points
behaves according to the power law
\begin{equation}\label{caa2}
M_{32}\sim\langle R_N\rangle^{d_{ip}}\sim N^{\nu_3d_{ip}}\>,
\end{equation}
where $\langle R_N\rangle\sim N^{\nu_3}$ is the mean end-to-end
distance of  the longer polymer chain. From the latter scaling
form, we find the following phenomenological formula  for the
contact critical exponent
\begin{equation}\label{foph}
\varphi_{32}^{_{CA}}=\left(1+{\nu_3\over\nu_2}\right)-\nu_3d_f^{3D}\>,
\end{equation}
whose  numerical values, evaluated from the attainable values for
$\nu_3$ and $\nu_2$ on SG fractals, are listed in table
\ref{tabela2}. Comparing the  values for $\varphi_{32}$ predicted
by the obtained CA formula with  our convincing set of MCRG data
(see figure~\ref{fig3}), we can perceive that in the first part of
the examined region ($2\le b\le40$), the CA formula displays
satisfactory agreement with the corresponding MCRG values, and,
for example, for $b=10$ fractal the two data (CA and MCRG) have
almost identical numerical values. For $b<10$, the CA formula
gives some smaller values for $\varphi_{32}$ than the MCRG method
(at most 9\%, for $b=2$), while for  $b>10$, the CA predicted
values get to be larger than the corresponding MCRG findings, and,
when $b$ increases, a departure of the two sets of data becomes
unambiguous (for $b=40$, the deviation is 57\%). Anyway, from both
the MCRG method and the CA formula (\ref{foph}), follow that
$\varphi_{32}$, in the studied region, is a monotonically
decreasing function of $b$.

Besides of  $\varphi_{32}$, we may define the   critical exponent
$\varphi_{33}$ which describes the number of polymer-polymer
contacts, when both polymers are floating chains in the 3D SG
fractal container (that is, when both polymers are of the type
$P_3$), and $\varphi_{22}$ when both polymers are adsorbed by 2D
SG surface (both of the type $P_2$). For these exponents the CA
arguments give
\begin{equation}\label{ca33}
\varphi_{33}^{_{CA}}=2-\nu_3d_f^{3D}\>,
\end{equation}
\begin{equation}\label{ca22}
\varphi_{22}^{_{CA}}=2-\nu_2d_f^{2D}\>.
\end{equation}
A simple comparison of the data,  from table~\ref{tabela2},
reveals that  the following relationship holds
\begin{equation}\label{nej}
\varphi_{32}<\varphi_{33}<\varphi_{22}\>,
\end{equation}
for each particular  $b$, in the range $2\le b\le 40$. The first
inequality means that the number of polymer contacts increases
when both polymers being  situated in the bulk of 3D SG fractal,
because  of growing  fractal dimension of a polymer chain
 when it makes a transition
from the 2D SG surface  to the bulk of 3D SG fractal
($d_f^{P_3}=1/\nu_3>d_f^{P_2}=1/\nu_2$). On the other hand, in the
case when both polymers are adsorbed (situated on 2D SG fractal),
the fractal dimension of the embedded space decreases
$d_f^{2D}<d_f^{3D}$, and a  chance for two polymers to make a
contact increases, implying the second inequality in (\ref{nej}).

The  relation between $\varphi_{32}$ and $\varphi_{22}$, can be
tested on the reliable sets of MCRG values for $\varphi_{32}$
(calculated in this work) and $\varphi_{22}$ (calculated
previously \cite{z11}  in the range $2\le b\le 100$). The values
for $\varphi_{22}$, being always larger than the corresponding CA
values proposed by (\ref{ca22}), display a monotonic decreasing
from $\varphi_{22}(b=2)=0.7493$ up to $\varphi_{22}(b=100)=0.6735$
\cite{z11}. Their comparison with $\varphi_{32}$ (from
table~\ref{tabela1}) shows that the relation
$\varphi_{32}<\varphi_{22}$ is always satisfied. The  relations
between $\varphi_{33}$ and the other two contact critical
exponents ($\varphi_{32}$ and $\varphi_{22}$) can be numerically
proved only for $b=2$ fractal, where the RG value  is known
$\varphi_{33}(b=2)=0.6635$ \cite{ksjstat}, and from which  the CA
value 0.6516 deviates only 1.8\%.   To complete the inspection of
relation (\ref{nej}) and  test CA formula (\ref{ca33}) for larger
$b$, one needs to calculate  $\varphi_{33}$ values for $b>2$,
which appears to be much complicated task than the calculation of
$\varphi_{32}$, and may be a topic for a future study.

Finally, we discuss the possible behaviour of the contact critical
exponents in the fractal-to-Euclidean crossover region, that is,
in the limit $b\to\infty$ ($d_f^{3D}\to3$, $d_f^{2D}\to2$).  For a
SAW situated in Euclidean spaces, the end-to-end distance critical
exponent in $d=2$ takes the exact value $\nu_2^E=3/4$
\cite{Nienhuis}, that was numerically confirmed \cite{jensen} for
the triangular lattice (to which the 2D SG fractals approach, when
$b\to\infty$), whereas in $d=3$, at present, the most accurate
evaluates are $\nu_3^E=0.5874\pm0.0002$ \cite{n31} and
$\nu_3^E=0.58765\pm0.00020$ \cite{n32}. Consequently, from the
formulas ({\ref{foph})--(\ref{ca22}), we compute the Euclidean
values: $\varphi_{32}^{E}\approx 0$, $\varphi_{33}^{E}\approx 1/5$
and $\varphi_{22}^{E}= 1/2$. In the case of 2D SG fractals (when
both polymers are of the type $P_2$), using the finite-size
scaling arguments it was argued \cite{ksjstat}, that in the limit
$b\to\infty$, the asymptotic behaviour of $\varphi_{22}$ is
described by the relation $\varphi_{22}\simeq 2-\nu_2d_f^{2D}$,
which coincides with the CA formula (\ref{ca22}). Knowing, from
the finite-size scaling analysis \cite{Dhar2}, that $\nu_2$ for
very large $b$ goes, from below, to the Euclidean value $3/4$,
from the latter asymptotic relation follows that $\varphi_{22}$
approaches, from above, the Euclidean value $\varphi_{22}^E=1/2$,
when $b\to\infty$.

As regards the obtained MCRG results for $\varphi_{32}$, it is
hard to say what happens beyond $b=40$, and specially, to
establish whether the critical exponent $\varphi_{32}$, in the
limit $b\to\infty$, goes to the zero Euclidean value, or some
non-Euclidean value (see figure \ref{fig3}). To  answer this
question properly, another method may be needed, for instance, the
finite-size scaling approach that has been applied in the case of
2D SG fractals \cite{ksjstat,Dhar2}. Here, we can only discuss
possible predictions, about the limiting value for $\varphi_{32}$,
that follow from the corresponding CA formula, for which  we need
to know the asymptotic behaviour  of $\nu_3$. Unfortunately, up
today, the behaviour of $\nu_3$, in the fractal-to-Euclidean
crossover region $b\to\infty$, exists as an unsolved problem. At
this moment, we may say that if $\nu_3\to\nu_3^E$ (which means,
see data from table~\ref{tabela2}, that $\nu_3$ is a non-monotonic
function of $b$), then ({\ref{foph}) gives $\varphi_{32}\to
\varphi_{32}^{E}\approx0$. On the contrary, if $\nu_3$ tends to
some non-Euclidean value less than $\nu_3^E$, then $\varphi_{32}$
will go to some value larger than the Euclidean value
$\varphi_{32}^E$. The possibility that $\nu_3$, for $b\to\infty$,
tends to some value larger than $\nu_3^E$ should imply a negative
value of $\varphi_{32}$ (and consequently, in this case, we cannot
exclude a situation in which the studied transition turns into
first order, with increasing $b$). Similar conclusions may be
deduced about the limiting value for the contact critical exponent
$\varphi_{33}$. Particulary, from (\ref{ca33}) follows: if
$\nu_3\to\nu_3^E$ then
$\varphi_{33}\to\varphi_{33}^{E}\approx1/5$, or,  if $\nu_3$
approaches the non-Euclidean value,  then $\varphi_{33}$ also
approaches a non-Euclidean value. From the exposed analysis we may
infer  that the behaviour of the contact critical exponents, in
the fractal-to-Euclidean crossover region, is closely related to
the corresponding behaviour of the end-to-end distance critical
exponents.

\section{Summary}
\label{cetiri}

In this paper we have studied  the  two interacting linear polymer
chains, modelled by two mutually crossing SAWs, situated on
fractal structures represented by the three-dimensional (3D)
Sierpinski gasket (SG) family of fractals. We take on that the
first polymer ($P_3$) is a floating chain in the bulk of 3D SG
fractal, while the second polymer chain ($P_2$) is adsorbed onto
one of the four boundaries  of the 3D SG fractal, which appears to
be 2D SG fractal. Specifically, we have calculated  the contact
critical exponent $\varphi_{32}$, associated with the number of
monomer-monomer contacts between
 polymers $P_3$ and $P_2$.

By applying the  renormalization group (RG) method, we have
calculated the exact value of the critical exponent
$\varphi_{32}$, for the first member $b=2$ of 3D SG fractal
family. The specific accomplishment in the course of this work is
the calculation of a long sequence of values of $\varphi_{32}$,
for $2\le b\le 40$, obtained by applying the Monte Carlo
renormalization group (MCRG)  method. Our results demonstrate that
$\varphi_{32}$, for the studied values of $b$, monotonically
decrease with $b$, and, in  analogy with  the behaviour of
$\varphi_{22}$ (that governs the number of contacts between two
adsorbed polymers) it seems that for $b>40$ the critical exponent
$\varphi_{32}$ continues decreasing, and in the limit $b\to\infty$
tends to the Euclidean value $\varphi_{32}^E\approx0$. Finally,
using the codimension additivity (CA) arguments we have proposed
the phenomenological formulae for the considered contact critical
exponents, and we have tested their predictions on the obtained
sets of data. We find that, for fractals labelled by smaller
values of $b$, the CA proposals give satisfactory agreement with
existing convincing results.

On the exposed grounds of the presented investigation we may
conclude that the set of obtained results of the studied problem
has been significantly extended. We have demonstrated that the
statistics of two crosslinked polymer chains on the  family of 3D
SG fractals can be rewardingly studied by the MCRG method. In
particular, the MCRG study of the contact critical exponents
revealed their interesting behaviour as the functions of fractal
scaling parameter $b$, making a step forward  to prescribe the
behaviour of SAW critical exponents in the fractal-to-Euclidean
crossover region. As a further investigation one may attempt to
extend our study for calculating the contact critical exponent
$\varphi_{33}$, when both polymers are floating chains in the bulk
of  3D SG fractal. In addition, to make the studied model more
realistic (but more difficult to study), it can be supplemented by
 additional interactions. For instance, one can introduce the interactions
between the bulk floating chain and the sites of adsorbing surface
(to promote the adsorbed phase of floating chain), as well as the
intra-chain interactions (to promote a collapsing phase of the
bulk floating chain), and then, in the space of interaction
parameters the phase diagram can be investigated, together with
new contact critical exponents at appropriate phase transition
fixed points.

\ack{The author thanks  S Elezovi\'c-Had\v zi\'c for critical
reading of the manuscript, and for useful discussions at the
beginning of this work. This paper has been done as a part of the
work within the project No.141020B funded by the Serbian Ministry
of Science and Protection of the Life Environment.}

\end{document}